\documentclass{ccjnl}
\usepackage{amssymb}
\usepackage{overpic}
\usepackage{algorithm}
\usepackage{algorithmic}
\usepackage{array}
\usepackage{graphicx}
\usepackage{color}
\usepackage{epstopdf}
\usepackage{acronym}
\usetikzlibrary{arrows}
\usetikzlibrary{decorations}

\definecolor{BLUE}{rgb}{0,0,1}

\acrodef{aoa}[AOA]{angle-of-arrival}
\acrodef{acf}[ACF]{autocorrelation function}
\acrodef{bcrb}[BCRB]{Bayesian Cram\'{e}r-Rao bound}
\acrodef{bp}[BP]{belief propagation}
\acrodef{cdi}[CDI]{cooperative dilution intensity}
\acrodef{cl}[CL]{cooperative localization}
\acrodef{cdf}[CDF]{cumulative distribution function}
\acrodef{crb}[CRB]{Cram\'{e}r-Rao bound}
\acrodef{crlb}[CRLB]{Cram\'{e}r-Rao lower bound}
\acrodef{dof}[DoF]{degree of freedom}
\acrodef{dct}[DCT]{discrete cosine transform}
\acrodef{dpeb}[DPEB]{directional position error bound}
\acrodef{fim}[FIM]{Fisher information matrix}
\acrodef{efim}[EFIM]{equivalent Fisher information matrix}
\acrodef{ici}[ICI]{information coupling intensity}
\acrodef{icrb}[ICRB]{inverse CRB}
\acrodef{iid}[i.i.d.]{independently and identically distributed}
\acrodef{isac}[ISAC]{integrated sensing and communication}
\acrodef{mse}[MSE]{mean-squared error}
\acrodef{pdf}[PDF]{probability density function}
\acrodef{peb}[PEB]{position error bound}
\acrodef{speb}[SPEB]{squared position error bound}
\acrodef{pll}[PLL]{phase-locked loop}
\acrodef{rbs}[RBS]{reference broadcast synchronization}
\acrodef{rhs}[RHS]{right hand side}
\acrodef{rii}[RII]{ranging information intensity}
\acrodef{rss}[RSS]{received signal strength}
\acrodef{rc}[RC]{ranging coefficient}
\acrodef{speb}[SPEB]{squared position error bound}
\acrodef{toa}[TOA]{time-of-arrival}
\acrodef{tdoa}[TDOA]{time-difference-of-arrival}
\acrodef{tpsn}[TPSN]{time synchronization protocol for sensor network}
\acrodef{vmp}[VMP]{variational message passing}
\acrodef{wsn}[WSN]{wireless sensor network}
\acrodef{efim}[EFIM]{equivalent Fisher information matrix}
\acrodef{dio}[DIO]{distance-information-only}
\acrodef{aio}[AIO]{angle-information-only}
\acrodef{saaf}[SAAF]{squared array aperture function}
\acrodef{snc}[S\&C]{sensing and communications}
\acrodef{uoa}[UOA]{uniformly oriented array}
\acrodef{rgg}[RGG]{random geometric graph}
\acrodef{rms}[RMS]{root-mean-square}
\acrodef{snr}[SNR]{signal-to-noise ratio}
\acrodef{eoc}[EoC]{efficiency of cooperation}
\acrodef{npi}[NPI]{nominal position information}
\acrodef{gnss}[GNSS]{global navigation satellite system}
\acrodef{mimo}[MIMO]{multiple-input multiple-output}
\acrodef{mcs}[MCS]{minimally constrained system}
\acrodef{zzb}[ZZB]{Ziv-Zakai bound}
\acrodef{wwb}[WWB]{Weiss-Weinstein lower bound}
\acrodef{nlos}[NLOS]{non-light-of-sight}
\acrodef{mmse}[MMSE]{minimum mean squared error}
\acrodef{uav}[UAV]{unmanned aerial vehicle}
\acrodef{ppp}[PPP]{Poisson point process}
\acrodef{bpp}[BPP]{binomial point process}
\acrodef{cln}[CLN]{cooperative location-aware network}
\acrodef{pdr}[PDR]{pedestrian dead reckoning}
\acrodef{ml}[ML]{maximum likelihood}
\acrodef{map}[MAP]{maximum \textit{a posteriori}}
\acrodef{ofdm}[OFDM]{orthogonal frequency division multiplexing}
\acrodef{cp}[CP]{cyclic prefix}
\acrodef{pacs}[P-ACS]{periodic autocorrelation sequence}
\acrodef{eisl}[EISL]{expected integrated sidelobe level}
\acrodef{dft}[DFT]{discrete Fourier transform}
\acrodef{6g}[6G]{sixth-generation}
\acrodef{drt}[DRT]{deterministic-random tradeoff}
\acrodef{awgn}[AWGN]{additive white Gaussian noise}
\acrodef{psk}[PSK]{phase shift keying}
\acrodef{qam}[QAM]{quadrature amplitude modulation}
\acrodef{ba}[BA]{Blahut-Arimoto}

\acrodefplural{dof}[DoFs]{degrees of freedom}

\usepackage{graphicx}
\usepackage{amsfonts}
\usepackage{amssymb}
\usepackage{amsmath}
\usepackage{color}
\usepackage{mathrsfs}
\usepackage{upgreek}
\usepackage{algorithm}
\usepackage{algorithmic}
\usepackage{winsnotation}
\usepackage{footnote}
\usepackage{cuted}

\title{Input Distribution Design for Ranging-Oriented OFDM-ISAC Systems Under Frequency-Selective Fading}
\author{Weijiang Zhao\inst{1}, Yifeng Xiong\inst{1,*}\corinfo{yifengxiong@bupt.edu.cn}}
\receiveddate{}
\reviseddate{}
\Editor{}

\address[1]{School of Information and Communication Engineering, Beijing University of Posts and Telecommunications, Beijing 100876, China}

\begin{document}
\maketitle

\begin{abstract}
The implementation of the \ac{isac} feature in \ac{6g} networks is most likely to be based on the framework of \ac{ofdm}. Input distribution design, or constellation design, is a crucial technique in \ac{ofdm}-\ac{isac} systems enabling a favorable balance between communication rate and sensing performance. In this treatise, we propose a computationally efficient input distribution design approach for \ac{ofdm}-\ac{isac} under frequency-selective channels, following the theoretical framework of capacity distortion. We highlight that under practical sensing constraints, the optimal strategy is to treat the kurtosis of constellations as a resource, and allocate it appropriately over subcarriers.
\keywords{OFDM; ISAC; capacity-distortion; deterministic-random tradeoff; constellation design}
\end{abstract}

\section{introduction}\label{sec:intro}
\Acf{isac} is deemed as one of the six major usage scenarios in \acf{6g} wireless networks \cite{ITU2023}, with applications spanning from body-area networks to low-altitude economy and satellite networks \cite{Chafii2023CST,saad2019vision,9737357,has_isac}. The main idea of \ac{isac} is to share wireless resources and infrastructures across communication and sensing functionalities in a single network. Among numerous technical challenges, one of the most timely and critical ones is to determine the waveforms and signals used in \ac{isac} systems, which are capable of supporting both target sensing and information delivery services over the same channel in a resource-efficient manner, as indicated by recent meetings of the 3rd-Generation Partnership Project (3GPP) \cite{ran_meeting}. Considering the architectural stability and implementation cost, the most likely evolution path forward in \ac{6g} networks is to design waveforms under the framework of \acf{ofdm}.

In particular, \ac{ofdm}-based design falls into the category of communication-centric \ac{isac} systems \cite{10012421,9921271,liao2024pulse}, which aim for reusing communication waveforms for sensing, with slight modifications. To enhance the sensing capability of these systems, a promising approach is to jointly use reference signals and data payloads for sensing \cite{liao2024pulse,zhang2023input,Keskin2024fundamental}. Against this background, one of the fundamental performance-limiting mechanisms is known to be the \textit{\ac{drt}} \cite{10147248,10471902,liu2023deterministic}. Specifically, \ac{drt} refers to the fact that the sensing functionality achieves its optimal performance at certain waveforms that are carefully designed, and hence deterministic, whereas the intrinsic randomness in the data payloads causes degradations in sensing performance. Given the performance metrics of sensing and communication, it is thus essential to develop techniques adjusting the \ac{drt} in a Pareto-optimal manner.

\ac{drt}-adjusting techniques have become a topic of active research \cite{du2024reshaping,sturm2011waveform,9005192,9724170,9109735,10463758,9359665,10264814,10638525}. For the specific task of target detection and ranging, the most widely used sensing performance metric is the sidelobe levels of the autocorrelation function (or more generally, the ambiguity function) of the transmitted \ac{isac} signal \cite{9724170,iceberg,preprint_opt_ofdm}. To elaborate, sensing signals having lower sidelobe levels are more favorable, since weaker targets would then be more likely to be resolvable from stronger targets in their vicinity. For \ac{isac} systems, the expected sidelobe level is a more suitable metric due to the randomness of the signals. In this context, the analytical expression of the expected sidelobe levels is given in \cite{preprint_opt_ofdm}, in which the most relevant take-home message is that the expected sidelobe level is related to the kurtosis of constellations. Kurtosis is a statistic that takes larger values for \ac{qam} constellations, whereas it takes smaller values for \ac{psk} constellations. This suggests that the value of kurtosis is also related to the achievable communication rate. Following this line of reasoning, in \cite{zhang2023input}, it is confirmed that \ac{psk} constellations yield substantially lower sidelobe levels compared to their \ac{qam} counterparts. For \ac{awgn} channels, \cite{du2024reshaping} proposed a probabilistic constellation shaping approach that adjusts the value of kurtosis to strike a beneficial balance between the expected sidelobe level and the achievable communication rate. 

In its nature, probabilistic constellation shaping designs the input distribution of the \ac{isac} channel, which can be formulated more systematically under the capacity-distortion framework \cite{10471902,9785593,10153971,9787809}. For \ac{ofdm} systems undergoing the simplistic \ac{awgn} channels, the optimal strategy is known to be transmitting \ac{iid} symbols over subcarriers, which admits computationally efficient design approaches that belong to the class of modified \ac{ba} algorithms \cite{9785593}. Nevertheless, practical \ac{ofdm} systems typically suffer from frequency-selective fading that complicates the problem. In particular, the modified \ac{ba} algorithm involves high-dimensional numerical integrations for vector channels, including frequency-selective channels, and thus would become computationally prohibitive for \ac{ofdm} systems containing a large number of subcarriers.

In this treatise, we propose a computationally efficient approach to design input distributions for frequency-selective \ac{isac} channels in \ac{ofdm} systems. Our main contributions are summarized as follows.
\begin{itemize}
\item We formulate the input distribution design problem as a mutual information maximization problem under power budget and \ac{eisl} constraints;
\item Under the \ac{ba} framework, we provide an efficient algorithm framework by showing that the optimal input distribution is factorizable across subcarriers;
\item To further reduce the complexity and enable real-time design, we propose a gradient projection-based approach, in which the gradients and the projection operators are computed using only closed-form expressions and one-dimensional searches;
\item We analyse the characteristics of the input distributions obtained by the proposed method. In particular, using numerical computations, we demonstrate that uniform power allocation is more favorable when the \ac{eisl} constraint is stringent, and hence the optimal strategy is to allocate kurtosis over subcarriers.
\end{itemize}

The rest of this treatise is organized as follows. Section \ref{sec:model} introduces the model of the \ac{ofdm}-\ac{isac} system, as well as the associated performance metrics. Section \ref{sec:framework} formulates the input distribution design problem and presents the computationally efficient algorithm framework based on factorization. Section \ref{sec:approx} elaborates on the real-time design approach based on gradient projection, with an approximation to the mutual information. Section \ref{sec:discuss} discusses the properties of the input distributions, using both analytical and numerical arguments. Section \ref{sec:numerical} provides numerical results to evaluate the performance of the proposed method. Finally, We conclude the treatise in Section \ref{sec:conclusion}.

\subsection*{Notations}
Throughout this treatise, $\rv{a}$, $\RV{a}$, $\RM{A}$, and $\RS{A}$ represent random variables (scalars), random vectors, random matrices and random sets, respectively; Their realizations, or the corresponding deterministic quantities, are denoted by $a$, $\V{a}$, $\M{A}$, and $\Set{A}$, respectively. The $m$-by-$n$ matrix of zeros (resp. ones) is denoted by $\M{0}_{m\times n}$ (resp. $\M{1}_{m\times n}$). The $m$-dimensional vector of zeros (resp. ones) is denoted by $\V{0}_{m}$ (resp. $\M{1}_{m}$). The $m$-by-$m$ identity matrix is denoted by $\M{I}_{m}$. These subscripts are omitted if they are clear from the context. The notation $[\cdot]_{i,j}$ denotes the $(i,j)$-th entry of its argument. The Hadamard product between matrices (or vectors) $\M{A}$ and $\M{B}$ is denoted by $\M{A}\odot\M{B}$. $\|\V{x}\|_p$ denotes the $l_p$ norm, which represents the $l_2$ norm by default when the subscript is omitted. $|\V{x}|^2$ denotes the vector containing the entrywise squared magnitudes of $\V{x}$.

\section{system model}\label{sec:model}
We consider an \ac{ofdm}-\ac{isac} system with $N$ subcarriers, and focus on the transmission of a single data stream. Assuming that the \ac{cp} is sufficiently long, the frequency-domain communication channel may be expressed as follows
\begin{equation}
\rv{y}_i = h_i\rv{x}_i + \rv{n}_i,\qquad i=1\dotsc N,
\end{equation}
where $\rv{n}_i$ denotes the additive noise on the $i$-th subcarrier modelled as circularly symmetric complex Gaussian distributed random variable with unit variance, i.e., $\rv{n}_i\sim \mathcal{CN}(0,1)$, while $\rv{x}_i$ and $h_i$ denote the communication symbol and the complex channel gain on the $i$-th subcarrier, respectively. We assume that the communication channel is quasi-static, and that all $h_i$'s are known due to channel estimation conducted prior to the transmission signal design. To facilitate further analysis, we denote
$$
\begin{aligned}
\V{h}&:=[h_1,~h_2,~\dotsc,~h_N]^{\rm T},\\
\RV{x}&:=[\rv{x}_1,~\rv{x}_2,~\dotsc,~\rv{x}_N]^{\rm T},\\
\RV{y}&:=[\rv{y}_1,~\rv{y}_2,~\dotsc,~\rv{y}_N]^{\rm T},\\
\RV{n}&:=[\rv{n}_1,~\rv{n}_2,~\dotsc,~\rv{n}_N]^{\rm T}.\\
\end{aligned}
$$
Using these notations, we obtain
\begin{equation}\label{conditional}
p_{\RV{y}|\RV{x}}(\V{y}|\V{x}) \propto \exp\left(-\|\RV{y}-\V{h}\odot \RV{x}\|_2^2\right).
\end{equation}

We consider the natural performance metric, namely the achievable rate, for the communications system. The optimal achievable rate is given by the channel capacity
\begin{equation}
\max_{p_{\RV{x}}(\V{x})}~~ I(\RV{x};\RV{y}),
\end{equation}
under the constraints of the power budget and sensing performance. As for the sensing performance metric, it is well-known that frequency-domain wireless resources are related to ranging accuracy. In light of this, we use the sidelobe level of the normalized \ac{pacs} of the time-domain transmitted signal to characterize the sensing performance. \endnote{In this treatise, we consider the normalized \ac{pacs} for the sake of simplicity. The unnormalized \ac{pacs} is given by $\tilde{\RV{r}_{\RV{x}}}=\sqrt{N}\M{F}^{{\rm{H}}}|\RV{x}|^2$.} In particular, we consider the \acf{eisl}
\begin{equation}
{\rm EISL}= \mathbb{E}\left\{\sum_{i=2}^N|[\RV{r}_{\RV{x}}]_i|^2\right\}
\end{equation}
introduced in \cite{preprint_opt_ofdm} to account for the randomness of the communication symbols, where $\RV{r}_{\RV{x}}$ denotes the vector corresponding to the \ac{pacs}, which can be expressed in terms of $\RV{x}$ as follows
\begin{equation}\label{pacs}
\RV{r}_{\RV{x}} = \M{F}^{\rm{H}}|\RV{x}|^2,
\end{equation}
with $\M{F}$ being the $N$-point \ac{dft} matrix given by
$$
\M{F}=\frac{1}{\sqrt{N}}\left[ {\begin{array}{*{20}{c}}
\omega_N^{0\cdot 0} & \omega_N^{0\cdot 1} & \dotsc & \omega_N^{0\cdot (N-1)}\\
\omega_N^{1\cdot 0} & \omega_N^{1\cdot 1} & \dotsc & \omega_N^{1\cdot (N-1)}\\
\vdots & \vdots & \ddots & \vdots \\
\omega_N^{(N-1)\cdot 0} & \omega_N^{(N-1)\cdot 1} & \dotsc & \omega_N^{(N-1)\cdot (N-1)}
\end{array}}\right]
$$
with $\omega_N=\exp(-2\pi i/N)$. A lower value of \ac{eisl} indicates that weaker targets are less likely to be overwhelmed by stronger targets, and hence the distance estimation accuracy can be more satisfactory. 

Another statistic that would play an important part in the subsequent analysis is the \textit{kurtosis} defined as
$$
\kappa_i = \frac{\mathbb{E}\{|\rv{x}_i|^4\}}{[\mathbb{E}\{|\rv{x}_i|^2\}]^2}.
$$
As will be clear in the following Sections, the \ac{eisl} can be expressed in terms of kurtoses.

\section{Optimization framework}\label{sec:framework}
We may now formulate our signal design problem as the maximization of the mutual information between $\RV{x}$ and $\RV{y}$, while satisfying the power budget constraint as well as the \ac{eisl} constraint, given by
\begin{subequations}\label{opt_original}
\begin{align}
\max_{p_{\RV{x}}(\V{x})}&~~I(\RV{x};\RV{y})\\
{\rm s.t.}&~~ \mathbb{E}\{\|\RV{x}\|_2^2\} = P, \\
&~~\mathbb{E}\left\{\sum_{i=2}^N|[\RV{r}_{\RV{x}}]_i|^2\right\} \leq D. \label{distortion_constraint}
\end{align}
\end{subequations}
Using \eqref{pacs} and the unitarity of the \ac{dft} matrix, we have the following result.
\begin{proposition}
Assuming that all $\rv{x}_i$'s are independent of each other, the \ac{eisl} can be expressed as
\begin{align}
{\rm EISL} &= \frac{N-1}{N}\mathbb{E}\{\|\RV{x}\|_4^4\} -\frac{1}{N}(\mathbb{E}\{\|\RV{x}\|_2^2\})^2 \nonumber\\
&\hspace{3mm}+\frac{1}{N}\sum_{i=1}^N (\mathbb{E}\{|\rv{x}_i|^2\})^2. \label{eisl_expression}
\end{align}
\begin{proof}
See Appendix \ref{app:proof_EISL}.
\end{proof}
\end{proposition}
The problem \eqref{opt_original} can now be simplified as
\begin{subequations}\label{opt_simplified}
\begin{align}
\max_{p_{\RV{x}}(\V{x})}&~~I(\RV{x};\RV{y})\\
{\rm s.t.}&~~ \mathbb{E}\{\|\RV{x}\|_2^2\} = P, \label{power_constraint2} \\
&~~\mathbb{E}\{\|\RV{x}\|_4^4\} + \frac{1}{N-1}\sum_{i=1}^N(\mathbb{E}\{|\rv{x}_i|^2\})^2 \nonumber\\
&~~\hspace{5mm}\leq \widetilde{D} + \frac{1}{N-1}P^2,\label{distortion_constraint2}
\end{align}
\end{subequations}
where $\widetilde{D}=\frac{ND}{N-1}$. We may further recast the problem as a double maximization given by
\begin{subequations}
\begin{align}
\max_{\V{p}}&~~R(\V{p})\\
{\rm s.t.}&~~\V{1}^{\rm T}\V{p}= P,\\
&~p_i\geq 0,~\forall i,
\end{align}
\end{subequations}
where the function $R(\V{p})$ is given by
\begin{subequations}\label{inner_opt}
\begin{align}
R(\V{p}) = \max_{p_{\RV{x}}(\V{x})}&~~I(\RV{x};\RV{y})\\
{\rm s.t.}&~~\mathbb{E}\{|\rv{x}_i|^2\}=p_i,\\
&~~\mathbb{E}\{\|\RV{x}\|_4^4\}\leq \widetilde{D}+\frac{P^2-\|\V{p}\|_2^2}{N-1}.
\end{align}
\end{subequations}
Later we shall see that the independence assumption of $\rv{x}_i$'s does not hurt the generality of the analysis.

In general, a cost-constrained mutual information maximization problem in the form of \eqref{inner_opt} can be solved by the modified \ac{ba} algorithm \cite{10153971}, which is a fixed-point iteration between the following two steps:
\begin{enumerate}
\item Update the trail \textit{a posterior} distribution $q(\V{x}|\V{y})$ according to
\begin{equation}\label{trial_ap}
q(\V{x}|\V{y}) \leftarrow \frac{r(\V{x}) p_{\RV{y}|\RV{x}}(\V{y}|\V{x})}{\int r(\V{x}) p_{\RV{y}|\RV{x}}(\V{y}|\V{x}) {\rm d}\V{x}};
\end{equation}
\item Update the trail input distribution $r(\V{x})$ according to
\begin{equation}\label{update_trial_input}
r(\V{x})\leftarrow \frac{e^{\int p_{\RV{y}|\RV{x}}(\V{y}|\V{x}) \log q(\V{x}|\V{y}){\rm d}\V{y}-\V{\lambda}^{\rm T}|\RV{x}|^2-\mu \|\V{x}\|_4^4}}{\int e^{\int p_{\RV{y}|\RV{x}}(\V{y}|\V{x}) \log q(\V{x}|\V{y}){\rm d}\V{y}-\V{\lambda}^{\rm T}|\RV{x}|^2-\mu \|\V{x}\|_4^4} {\rm d}\V{x}},
\end{equation}
where $\V{\lambda}$ and $\mu$ represent the dual variables corresponding to the constraints \eqref{power_constraint2} and \eqref{distortion_constraint2}, respectively.
\end{enumerate}
Besides the fixed-point iteration, this algorithm requires an initialization of the trail input distribution $r(\V{x})$ and an appropriate search (or optimization) algorithm finding the values of $\V{\lambda}$ and $\mu$ satisfying the constraints \eqref{power_constraint2} and \eqref{distortion_constraint2}.

Despite being conceptually viable, the modified \ac{ba} algorithm is not computationally feasible for practical \ac{ofdm} systems, since the update of the trial input distribution $r(\V{x})$ involves an integral in extremely high-dimensional spaces\endnote{Up to several thousands, depending on the number of subcarriers.}. In what follows, we will derive a low-complexity, yet exact, approach to compute this integral.

We commence from a simplification of the objective function. Indeed, we have
\begin{subequations}
\begin{align}
I(\RV{x};\RV{y})&=h(\RV{y})-h(\RV{y}|\RV{x})\\
&=h(\RV{y})-h(\RV{n})\\
&=h(\RV{y})-\sum_{i=1}^N h(\rv{n}_i)\\
&\leq \sum_{i=1}^N \left[h(\rv{y}_i)-h(\rv{n}_i) \right]\\
&= \sum_{i=1}^N I(\rv{y}_i;\rv{x}_i).
\end{align}
\end{subequations}
This implies that the capacity is always achieved when the input distribution admits a factorization of
\begin{equation}
p_{\V{x}}(\V{x}) = \prod_{i=1}^N p_{\rv{x}_i}(x_i).
\end{equation}
Therefore, we may also use the following factorized trial input distribution
\begin{equation}\label{factor_trial_input}
r(\V{x}) = \prod_{i=1}^N r_i(x_i).
\end{equation}
Substituting \eqref{factor_trial_input} into \eqref{trial_ap}, we obtain
\begin{equation}
q(\V{x}|\V{y}) \leftarrow \frac{\prod_{i=1}^N r_i(x_i) p_{\RV{y}|\RV{x}}(\V{y}|\V{x})}{\int \prod_{i=1}^N r_i(x_i) p_{\RV{y}|\RV{x}}(\V{y}|\V{x}) {\rm d}\V{x}}.
\end{equation}
Furthermore, from \eqref{conditional} we see that $p_{\V{y}|\V{x}}(\V{y}|\V{x})$ is also factorizable, which enables us to use the following factorized trial \textit{a posteriori} distribution
\begin{equation}\label{factor_ap}
q_i(x_i|y_i) \leftarrow \frac{r_i(x_i) p_{\rv{y}_i|\rv{x}_i}(y_i|x_i)}{\int r_i(x_i) p_{\rv{y}_i|\rv{x}_i}(y_i|x_i) {\rm d} x_i},
\end{equation}
with $q(\V{x}|\V{y})=\prod_{i=1}^N q_i(x_i|y_i)$. We can now insert \eqref{factor_ap} and \eqref{factor_trial_input} into \eqref{update_trial_input}, and obtain
\begin{equation}
r_i(x_i) \leftarrow \frac{e^{g_i(x_i)}}{\int e^{g_i(x_i)}{\rm d}x_i},
\end{equation}
where
\begin{align}\label{update_g}
g_i(x_i) &= \int p_{\rv{y}_i|\rv{x}_i}(y_i|x_i) \log q_i(x_i|y_i){\rm d}y_i \nonumber \\
&\hspace{3mm}- \lambda_i |x_i|^2 -\mu |x_i|^4.
\end{align}

\begin{breakablealgorithm}
\caption{Factorized Modified \ac{ba} Algorithm.}
\label{alg:fmba}
\begin{algorithmic}
\REQUIRE Initialization of $r_i(x_i)$, $i=1\dotsc N$;
\ENSURE $p_{\rv{x}_i}(x_i)$, $i=1\dotsc N$;
\REPEAT
\FOR{i=1:N}
\STATE $q_i(x_i|y_i) \leftarrow \frac{r_i(x_i) p_{\rv{y}_i|\rv{x}_i}(y_i|x_i)}{\int r_i(x_i) p_{\rv{y}_i|\rv{x}_i}(y_i|x_i) {\rm d}x_i}$;
\STATE $g_i(x_i) \leftarrow \int p_{\rv{y}_i|\rv{x}_i}(y_i|x_i) \log q_i(x_i|y_i){\rm d}y_i- \lambda_i |x_i|^2 -\mu |x_i|^4$;
\STATE $r_i(x_i) \leftarrow \frac{e^{g_i(x_i)}}{\int e^{g_i(x_i)}{\rm d}x_i}$;
\ENDFOR
\UNTIL{Convergence condition is met}
\RETURN $p_{\rv{x}_i}(x_i)= r_i(x_i),~\forall i = 1\dotsc N$;
\end{algorithmic}
\end{breakablealgorithm}

The improved algorithm is summarized in Algorithm 1. Compared to the conventional modified \ac{ba} algorithm, the computational complexity is greatly reduced, since the integrals involved in the iterations are at most four-dimensional\endnote{Both $x_i$ and $y_i$ are complex scalars and hence have two real dimensions.}. 

Nevertheless, implementing the algorithm using straightforward numerical integrations can still be computationally prohibitive for real-time transmission signal design, not to mention that the dual variables $\V{\lambda}$ and $\mu$ have to be searched outside the loop. In the next section, we discuss some practical implementation strategies.

\section{Approximate Optimization Algorithm}\label{sec:approx}
In this section, we provide some implementation strategies facilitating real-time signal design.

\subsection{Approximating the Mutual Information}
First, we note that there are mainly two drawbacks regarding the computational complexity of Algorithm 1: 
\begin{enumerate}
\item Given the values of $\V{\lambda}$ and $\mu$, the evaluation of the objective function involves multiple complicated numerical integrations (computed once per inner iteration); 
\item The search of $\V{\lambda}$ and $\mu$ is not efficient since the gradient (as well as higher-order quantities) is unknown.
\end{enumerate}
Naturally, we would like to construct algorithms that are capable of exploiting the gradient information, and preferably of reducing the complexity of numerical integration.

To this end, let us reformulate the optimization problem as follows
\begin{subequations}\label{problem_reformulated}
\begin{align}
\max_{\V{p},\V{d}}&~~\sum_{i=1}^N R_i(p_i,d_i),\\
{\rm s.t.}&~~\V{1}^{\rm T}\V{p}= P,\\
&~~\V{1}^{\rm T}\V{d}+\frac{N\|\V{p}\|_2^2}{N-1} \leq \frac{ND+P^2}{N-1},\\
&~~\V{p}\succeq \V{0},~\V{d}\succeq \V{0},
\end{align}
\end{subequations}
where
\begin{subequations}\label{rate_single_sc}
\begin{align}
R_i(p_i,d_i) := \max_{p_{\rv{x}_i}(x_i)}&~~I(\rv{x}_i;h_i\rv{x}_i+\rv{n}_i)\\
{\rm s.t.}&~~\mathbb{E}\{|\rv{x}_i|^2\}=p_i,\\
&~~\mathbb{E}\{|\rv{x}_i|^4\}=d_i+p_i^2,
\end{align}
\end{subequations}
with $p_i$ and $d_i$ being the constraints imposed on the second-order moment and the excess fourth-order moment, respectively. Next, note that
$$
I(\rv{x}_i;h_i\rv{x}_i+\rv{n}_i) = h(h_i\rv{x}_i+\rv{n}_i)-h(\rv{n}_i),
$$
where $h(\cdot)$ denotes the differential entropy\endnote{In this treatise, we derive the differential entropy using the natural logarithm for the sake of notational simplicity. In the numerical examples, we present the results in terms of bits via change of base.} of its argument, and that $h(\rv{n}_i)$ is a constant with respect to $p_{\rv{x}_i}(x_i)$, it suffices to replace $R_i(p_i,d_i)$ with $\tilde{R}_i(p_i,d_i)$ given by
\begin{subequations}\label{entropy_single_sc}
\begin{align}
\tilde{R}_i(p_i,d_i) := \max_{p_{\rv{x}_i}(x_i)}&~~h(h_i\rv{x}_i+\rv{n}_i)\\
{\rm s.t.}&~~\mathbb{E}\{|\rv{x}_i|^2\}=p_i,\\
&~~\mathbb{E}\{|\rv{x}_i|^4\}=d_i+p_i^2.
\end{align}
\end{subequations}

Now, using the following expressions of the moments
$$
\begin{aligned}
\mathbb{E}\{|h_i\rv{x}_i+\rv{n}_i|^2\} &= |h_i|^2\mathbb{E}\{|\rv{x}_i|^2\} +1,\\
\mathbb{E}\{|h_i \rv{x}_i+\rv{n}_i|^4\}&= |h_i|^4\mathbb{E}\{|\rv{x}_i|^4\}+4|h_i|^2\mathbb{E}\{|\rv{x}_i|^2\}+2,
\end{aligned}
$$
we obtain the following upper bound for the optimal objective function value of $\tilde{R}_i(p_i,d_i)$:
\begin{align}
&\tilde{R}_i(p_i,d_i)\leq \overline{R}_i(p_i,d_i)\nonumber \\
&=\max_{p_{\rv{z}}(z)}~~h(\rv{z})\nonumber\\
&\hspace{5mm}{\rm s.t.}~\mathbb{E}\{|\rv{z}|^4\}=|h_i|^4(d_i+p_i^2)+4|h_i|^2p_i+2,\nonumber\\
&\hspace{11mm}\mathbb{E}\{|\rv{z}|^2\}=|h_i|^2 p_i+1. \label{upper_bound}
\end{align}
This upper bound is tight when the solution to the following deconvolution equation yields a legitimate probability density function:
\begin{equation}\label{deconvolution}
\int p_{h_i\rv{x}_i}(h_ix_i) p_{\rv{n}_i}(z-h_ix_i){\rm d}x_i = p_{\rv{z}}^{\rm opt}(z),
\end{equation}
where $p_{\rv{z}}^{\rm opt}(z)$ denotes the optimal distribution obtained by solving \eqref{upper_bound}. When the bound is not tight, it can still serve as an approximation of the optimal achievable rate. In this case, an achievable strategy (that yields a suboptimal rate) is given by
$$
\begin{aligned}
p_{h_i\rv{x}_i}(h_ix_i)=\mathop{\arg\max}_{p_{\rv{z}}(z)}&~~h(\rv{z})\\
{\rm s.t.}&~~ \mathbb{E}\{|\rv{z}|^2\}=|h_i|^2p_i,\\
&~~\mathbb{E}\{|\rv{z}|^4\}=|h_i|^4(d_i+p_i^2).
\end{aligned}
$$
In either case, one has to solve an entropy maximization problem in the following unified form
\begin{subequations}\label{unified_maxent}
\begin{align}
\max_{p_{\rv{z}}(z)}&~~h(\rv{z})\\
{\rm s.t.}&~~ \mathbb{E}\{|\rv{z}|^2\}=M_2,\\
&~~\mathbb{E}\{|\rv{z}|^4\}=M_4.
\end{align}
\end{subequations}
The solution to such a problem is known to take the form of
\begin{equation}\label{maxent_dist}
p_{\rv{z}}(z) = \frac{1}{Z_0} \exp\left\{-c|z|^4+a|z|^2\right\},
\end{equation}
where $Z_0=\int \exp\left\{-c|z|^4+a|z|^2\right\}{\rm d}z$ is the normalization coefficient. Next, to enable efficient search of $\V{p}$ and $\V{d}$, we would like to compute the derivative of the optimal objective function value with respect to the natural parameters $a$ and $c$, and to express the constraints parameters $M_2$ and $M_4$ in terms of $a$ and $c$. In particular, we have the following result.

\begin{remark}[Extendibility of the Results]
We note that both the deconvolution method and the entropy maximization problem do not rely on the assumption that the noise is Gaussian distributed. To elaborate, the entropy maximization problem itself is independent of the noise, while the deconvolution problem is, by nature, applicable to additive noise following any distribution. One may consider even more realistic scenarios by accounting for time-varying channels or channel estimation errors. The proposed framework cannot be directly generalized to such scenarios and requires further investigations.
\end{remark}
\begin{proposition}\label{prop:diff_ent}
The differential entropy of $p_{\rv{z}}(z)$ in \eqref{maxent_dist} is given by
\begin{equation}
h(\rv{z}) = \ln \pi +\frac{1}{2} +\ln Z_1 -\frac{a}{4c}\cdot \frac{1}{Z_1}-\frac{a^2}{4c},
\end{equation}
where $Z_1$ is defined as
\begin{equation}
Z_1 =\sqrt{\frac{\pi}{4c}}e^{\frac{a^2}{4c}}\left[1+{\rm erf}\left(\frac{a}{2\sqrt{c}}\right)\right].
\end{equation}
The parameters $M_2$ and $M_4$ can be expressed in terms of the natural parameters $a$ and $c$ as
\begin{subequations}\label{moments_expression}
\begin{align}
M_2&=\frac{a}{2c}+\frac{1}{2cZ_1},\\
M_4&=\frac{1}{2c}(1+aM_2).
\end{align}
\end{subequations}
\begin{proof}
See Appendix \ref{app:proof_diff_ent}.
\end{proof}
\end{proposition}

\subsection{Gradient Projection Method}
Our next task is to compute the gradient of $h(\rv{z})$ with respect to the natural parameters $\V{\theta}=[a,c]^{\rm T}$, which is given in the following proposition.
\begin{proposition}\label{prop:grad_z}
The gradient of $h(\rv{z})$ with respect to $\V{\theta}$ is given by
\begin{equation}\label{grad_z}
\nabla_{\V{\theta}} h(\rv{z}) =\frac{1}{2cZ_1}\left[cM_4, -Z_1-\frac{1}{2}aM_4\right]^{\rm T}.
\end{equation}
\begin{proof}
See Appendix \ref{app:proof_grad_z}.
\end{proof}
\end{proposition}
Having the gradient computed, we may now solve the problem \eqref{problem_reformulated} using the gradient projection approach. The algorithm is summarized in Algorithm \ref{alg:gpa}, where $\alpha$ is a step size parameter, the notation $(x)_+$ denotes the rectified linear function $(x)_+=\max(x,0)$, and the function $\V{f}(\V{\theta})$ is defined as
\begin{equation}
\V{f}(\V{\theta})=\frac{1}{2\theta_2}[\theta_1+Z_1^{-1},1+\frac{\theta_1}{2\theta_2}(\theta_1+Z_1^{-1})]^{\rm T},  
\end{equation}
according to \eqref{moments_expression}, with $\V{f}^{-1}(\cdot)$ being its inverse function.

\begin{breakablealgorithm}
\caption{Gradient Projection Algorithm Solving \eqref{problem_reformulated}.}
\label{alg:gpa}
\begin{algorithmic}
\REQUIRE Initialization of $\V{p}^{(1|0)}$ and $\V{d}^{(1|0)}$;
\ENSURE Optimized $\V{p}$ and $\V{d}$;
\STATE $\ell=0$;
\REPEAT
\STATE $\ell\leftarrow \ell+1;$
\FOR{i=1:N}
\STATE $\V{\theta} \leftarrow \V{f}^{-1}\Big(|h_i|^2p_i^{(\ell|\ell-1)}+1,|h_i|^4\Big(d_i^{(\ell|\ell-1)}+(p_i^{(\ell|\ell-1)})^2\Big)+4|h_i|^2p_i^{(\ell|\ell-1)}+2\Big)$;
\STATE $\V{\theta}\leftarrow \V{\theta} + \alpha \nabla_{\V{\theta}} h(\rv{z})$;
\STATE $[p_{\rm temp},d_{\rm temp}]^{\rm T}\leftarrow \V{f}(\V{\theta})$;
\STATE $p_i^{(\ell|\ell)}\leftarrow \left(\frac{p_{\rm temp}-1}{|h_i|^2}\right)_+$;
\STATE $d_i^{(\ell|\ell)}\leftarrow \left(\frac{d_{\rm temp}-(p_{\rm temp}-1)^2-4(p_{\rm temp}-1)-2}{|h_i|^4}\right)_+$;
\ENDFOR
\STATE $[\V{p}^{(\ell+1|\ell)};\V{d}^{(\ell+1|\ell)}]\leftarrow {\rm Proj}(\V{p}^{(\ell|\ell)},\V{d}^{(\ell|\ell)})$;
\UNTIL{Convergence condition is met}
\RETURN $\V{p}= \V{p}^{(\ell+1|\ell)}$, $\V{d}= \V{d}^{(\ell+1|\ell)}$;
\end{algorithmic}
\end{breakablealgorithm}

Specifically, the operator ${\rm Proj}(\V{p}^{(\ell|\ell)},\V{d}^{(\ell|\ell)})$ projects the raw updates $\V{p}^{(\ell|\ell)}$ and $\V{d}^{(\ell|\ell)}$ to the feasible region, by solving the following problem
\begin{subequations}\label{problem_projection}
\begin{align}
\min_{\V{p},\V{d}}&~~\frac{1}{2}\left(\|\V{p}-\V{p}^{(\ell|\ell)}\|_2^2+\|\V{d}-\V{d}^{(\ell|\ell)}\|_2^2\right),\\
{\rm s.t.}&~~\V{1}^{\rm T}\V{p}= P,\label{original_condition_1}\\
&~~\V{1}^{\rm T}\V{d}+\frac{N\|\V{p}\|_2^2}{N-1} \leq \frac{ND+P^2}{N-1},\label{original_condition_2}\\
&~~\V{p}\succeq \V{0},~\V{d}\succeq \V{0}.\label{original_condition_3}
\end{align}
\end{subequations}
This is a convex quadratically constrained quadratic programming (QCQP) problem, which can be solved using off-the-shelf commercial solvers. Here, we provide a more efficient strategy by directly solving the Karush-Kuhn-Tucker (KKT) conditions, which are given by
\begin{subequations}
\begin{align}
(\V{p}-\V{p}^{(\ell|\ell)})+\lambda\V{1}+\frac{2N\mu}{N-1}\V{p}-\V{z}_{\V{p}}&=\V{0},\label{kkt_condition1}\\
(\V{d}-\V{d}^{(\ell|\ell)})+\mu\V{1}-\V{z}_{\V{d}}&=\V{0},\label{kkt_condition2}\\
\V{z}_{\V{p}}\succeq \V{0},~\V{z}_{\V{d}}\succeq \V{0},~\mu&\geq 0,\label{kkt_condition3}\\
\mu\left(\V{1}^{\rm T}\V{d}+\frac{N\|\V{p}\|_2^2}{N-1}-\frac{ND+P^2}{N-1}\right)&=0,\label{kkt_condition5}\\
\V{z}_{\V{p}}\odot \V{p}&=0,\label{kkt_condition6}\\
\V{z}_{\V{d}}\odot \V{d}&=0,\label{kkt_condition7}\\
\eqref{original_condition_1}-\eqref{original_condition_3}&,
\end{align}
\end{subequations}
where $\lambda$ and $\mu$ are the dual variables associated with the constraints \eqref{original_condition_1} and \eqref{original_condition_2}, respectively, whereas $\V{z}_{\V{p}}$ and $\V{z}_{\V{d}}$ are the dual variables associated with \eqref{original_condition_3}. Using \eqref{kkt_condition1} and \eqref{kkt_condition6} we obtain
\begin{equation}\label{p_condition}
p_i=\frac{N-1}{N-1+2N\mu}\left(p_i^{(\ell|\ell)}-\lambda\right)_+.
\end{equation}
Using \eqref{kkt_condition2} and \eqref{kkt_condition7} we have
\begin{equation}\label{d_condition}
d_i=(d_i^{(\ell|\ell)}-\mu)_+.
\end{equation}
Now, from \eqref{kkt_condition5} we may infer that $\mu=0$ holds when \eqref{original_condition_2} is inactive. Moreover, when $\mu=0$, the projection does not alter $\V{d}^{(\ell|\ell)}$, and hence $\V{d}^{(\ell|\ell)}$ itself would have already satisfied the constraint. In this case, one may obtain the value of $\V{p}$ by solving
$$
p_i=\left(p_i^{(\ell|\ell)}-\lambda\right)_+,~\V{1}^{\rm T}\V{p}=P.
$$
In general, given any $\mu\geq 0$, the value of $\lambda$ can be determined from \eqref{original_condition_1} and \eqref{p_condition}. For $\mu>0$, the equality of \eqref{original_condition_2} would also hold. Therefore, one may solve $\mu$ by first substituting \eqref{d_condition} into \eqref{original_condition_2}, and then performing an one-dimensional search for the positive real root of \eqref{original_condition_2} (as an equation).

\section{Discussions}\label{sec:discuss}
In this section, we discuss certain characteristics of the proposed methods.

\subsection{Characteristics of the Designed Input Distributions}
In this treatise, we have used the \ac{eisl} as the sensing performance metric. However, one might also be interested in the individual sidelobe levels, namely
$$
\mathbb{E}\{|[\RV{r}_{\rv{x}}]_i|^2\},~i=2,\dotsc,N.
$$
In the following proposition, we characterize the individual sidelobe levels.
\begin{proposition}
The expectation of the squared magnitudes of the \ac{pacs} is given by
\begin{equation}\label{exp_pacs}
\mathbb{E}\{|[\RV{r}_{\RV{x}}]_i|^2\}= |\V{f}_i^{\rm H}\V{p}|^2+\frac{1}{N}\sum_{j=1}^N (\kappa_j-1)p_j^2,
\end{equation}
where $\V{f}_i$ denotes the $i$-th column in $\M{F}$, $p_j=\mathbb{E}\{|\rv{x}_j|^2\}$, and $\kappa_j=\mathbb{E}\{|\rv{x}_j|^4\}/p_j^2$ denotes the kurtosis of $\rv{x}_j$. Especially, for $i=1$, we have
\begin{equation}
\mathbb{E}\{|[\RV{r}_{\rv{x}}]_1|^2\}= \frac{P^2}{N}+\frac{1}{N}\sum_{j=1}^N (\kappa_j-1)p_j^2.
\end{equation}
\begin{proof}
See Appendix \ref{app:proof_individual_sidelobe}.
\end{proof}
\end{proposition}

It is now clear that in order to reduce the sidelobe levels, one may: 1) reduce the kurtosis of $\rv{x}_i$'s, or 2) reduce the variance between the entries of $\V{p}$. To elaborate, all $\V{f}_i$'s ($i\neq 1$) are orthogonal to $\V{f}_1=\frac{1}{\sqrt{N}}\V{1}$, which implies that the term $|\V{f}_i^{\rm H}\V{p}|^2$ is zero when the entries of $\V{p}$ are identical. Especially, when these entries are identical, we have
$$
\mathbb{E}\{|[\RV{r}_{\rv{x}}]_i|^2\} = \frac{P^2}{N^3}\sum_{j=1}^N (\kappa_j-1)
$$
for $i\neq 1$, and 
$$
\mathbb{E}\{|[\RV{r}_{\rv{x}}]_1|^2\} = \frac{P^2}{N} + \frac{P^2}{N^3}\sum_{j=1}^N (\kappa_j-1).
$$
In this case, we see that the sidelobe levels are not directly determined by the kurtosis of individual subcarriers. Rather, the dependence is only conveyed by the average kurtosis given by
\begin{equation}
\bar{\kappa}=\frac{1}{N}\sum_{i=1}^N \kappa_i.
\end{equation}
This implies that when the entries of $\V{p}$ are identical, in order to control the individual sidelobes, it suffices to control the \ac{eisl}.

Naturally, one would ask when the power allocation over subcarriers is uniform. Apparently, when there is no \ac{eisl} constraint, the power allocation is in general not uniform, since the optimal strategy is the well-known ``water-filling'' power allocation with all kurtoses being equal to $2$ (corresponding to Gaussian distributions). The key fact utilized in the water-filling strategy is that the derivative of $h(\rv{z})$ with respect to $\mathbb{E}\{|\rv{z}|^2\}$ is non-negative, thus assigning more power to ``better'' subcarriers having larger SNRs would be more beneficial. However, in the presence of the \ac{eisl} constraint, the derivative is not necessarily non-negative. In particular, we are interested in the partial derivative $\frac{\partial h(\rv{z})}{\partial \mathbb{E}\{|\rv{z}|^2\}}$, the rate of change of $h(\rv{z})$ with respect to $\mathbb{E}\{|\rv{z}|^2\}$, given a specific value of the kurtosis $\kappa=\mathbb{E}\{|\rv{z}|^4\}/(\mathbb{E}\{|\rv{z}|^2\})^2$ is kept unchanged.

\begin{figure}[t]
    \centering
    \includegraphics[width=0.93\linewidth]{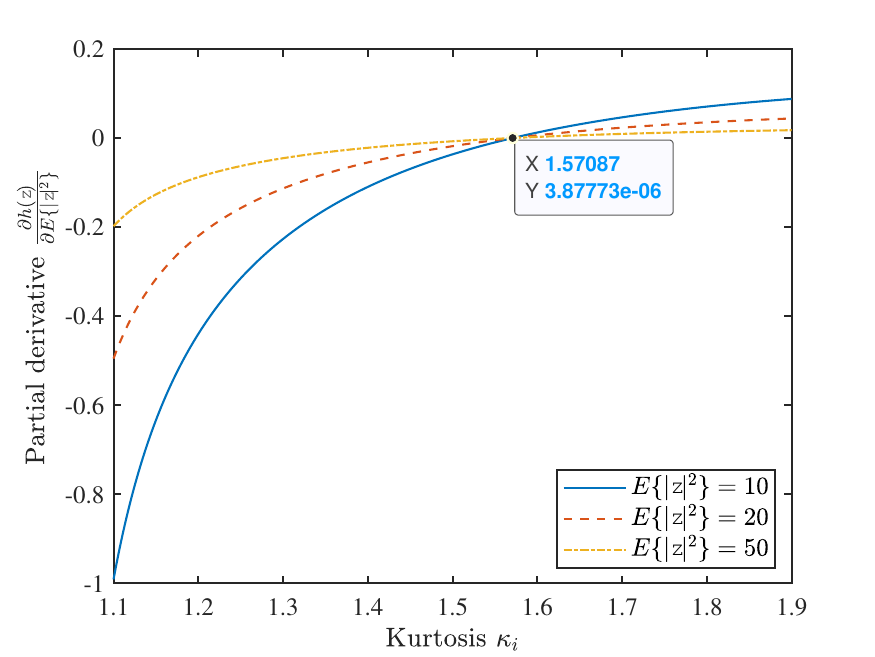}
    \caption{The numerically computed partial derivative $\frac{\partial h(\rv{z})}{\partial \mathbb{E}\{|\rv{z}|^2\}}$, which equals to zero when the kurtosis is around $1.57$.}
    \label{fig:partial_derivative}
\end{figure}

As is portrayed in Fig.\ref{fig:partial_derivative}, the (numerically computed) partial derivative is negative as long as the kurtosis is less than a specific value around $1.57$. Therefore, when the \ac{eisl} constraint is relatively stringent, assigning more power to better subcarriers is, counter-intuitively, not beneficial. In this case, the optimal strategy would be assigning power uniformly to all subcarriers, while assigning more kurtosis to subcarriers having higher SNRs (since this will not hurt the \ac{eisl}, or even individual sidelobe levels), as will be illustrated in Section \ref{sec:numerical}. 

\subsection{Complexity Analysis}
\begin{table}[htbp]
\caption{Complexity Analysis of the proposed methods and the conventional \ac{ba} algorithm.}
\label{tbl:complexity}
\centering
\begin{tabular}{c|c}
\hline
Method & Computational Complexity \\ \hline
Conventional Modified \ac{ba} & $O(N_{\rm GF}N_{\rm CBA}K^{4N})$ \\ \hline 
Factorized Modified \ac{ba} & $O(N_{\rm GF}N_{\rm FBA}NK^4)$ \\ \hline
Gradient Projection & $O(N_{\rm GP}NL\log_2 N)$ \\ \hline
\end{tabular}
\end{table}

In Table \ref{tbl:complexity}, we have compared the computational complexity of the proposed methods against the conventional modified \ac{ba} algorithm, where $K$ denotes the number of quadrature points on each dimension, $N_{\rm CBA}$ and $N_{\rm FBA}$ denote the number of \ac{ba} iterations for the conventional modified \ac{ba} and its factorized counterpart, respectively, while $N_{\rm GF}$ represents the number of search steps for the values of $\V{\lambda}$ and $\mu$ \endnote{One may rely on gradient-free optimization methods such as the Nelder-Mead method.}. For the gradient projection method, $N_{\rm GP}$ denotes the number of gradient projection steps, $L$ denotes the complexity of the one-dimensional searches, while the term $N\log_2 N$ corresponds to the water-filling subroutine involved in the projection step. In particular, the conventional modified \ac{ba} requires high-dimensional integration, which is computational prohibitive in the sense that the complexity is exponential in $N$. By factorizing the trial distributions, the factorized modified \ac{ba} improves the dependency on $N$ to the linear order. The gradient projection method further reduces the complexity by avoiding the complicated numerical integration and \ac{ba} iterations.

\section{numerical results}\label{sec:numerical}
In this section, we illustrate the analytical results and demonstrate the performance of the proposed method using numerical examples.

In particular, we consider an \ac{ofdm} system with $N=64$ subcarriers, undergoing a $4$-path Rician channel with a K-factor equals to $6$dB. If not otherwise stated, we assume that the SNR is $10$. We use the following \ac{eisl} constraint:
\begin{equation}
D = \frac{\zeta(N-1)P^2}{N^2}.
\end{equation}
When the power is uniformly assigned over subcarriers, we have $\zeta = \bar{\kappa}-1$. 

\begin{figure}[t]
    \centering
    \includegraphics[width=0.93\linewidth]{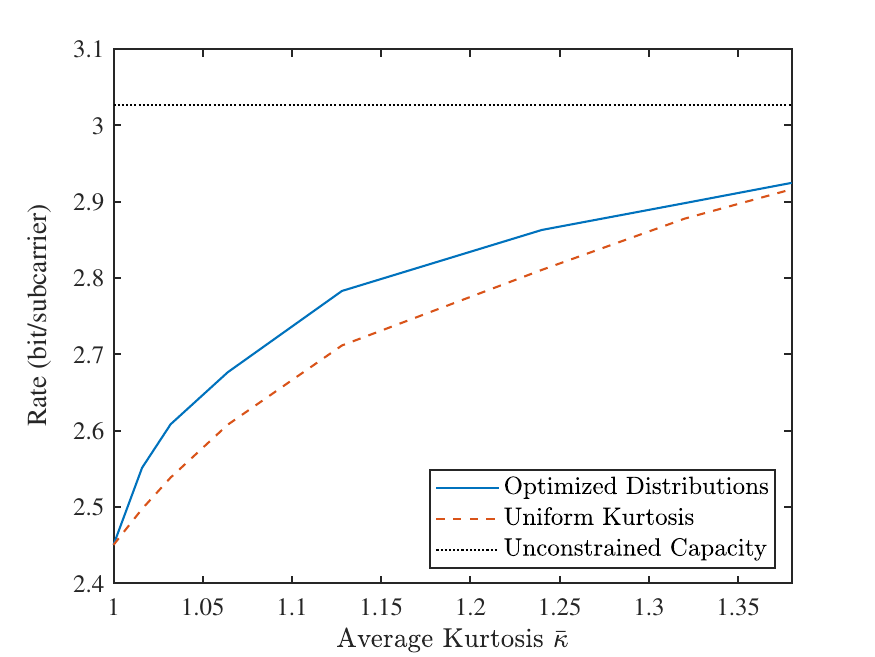}
    \caption{The rate-kurtosis tradeoff of the proposed method, compared to that of the uniform kurtosis allocation strategy.}
    \label{fig:rate_vs_kurtosis}
\end{figure}

\begin{figure}[t]
    \centering
    \includegraphics[width=0.93\linewidth]{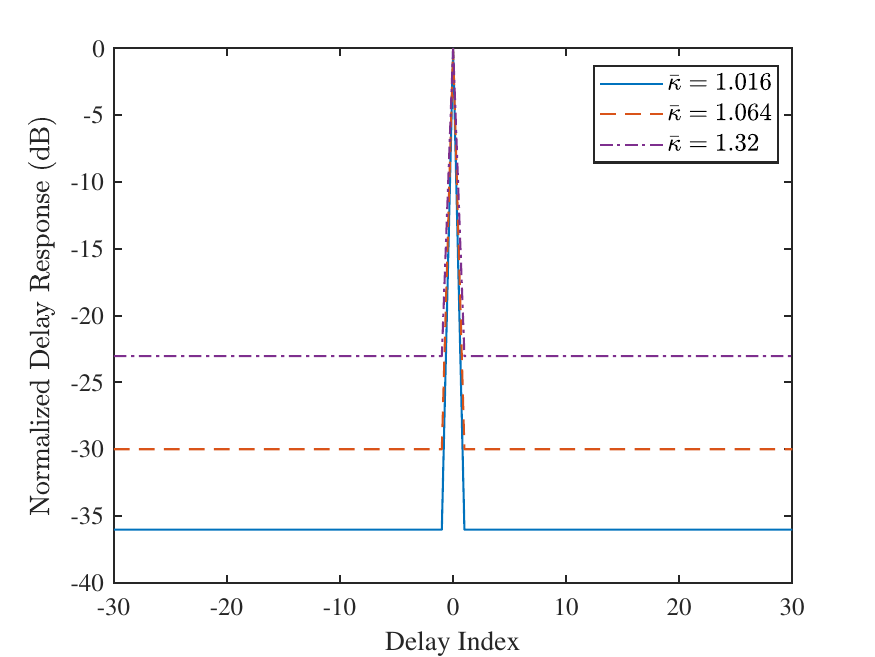}
    \caption{Graphical illustration of the normalized delay response $\mathbb{E}\{|[\RV{r}_{\RV{x}}]_i|^2\}/\mathbb{E}\{|[\RV{r}_{\RV{x}}]_1|^2\}$. The delay index is $i-1$.}
    \label{fig:pacs}
\end{figure}

Let us first demonstrate the tradeoff between the achievable rate and the average kurtosis $\bar{\kappa}$, as portrayed in Fig.~\ref{fig:rate_vs_kurtosis}. It turns out that within the considered interval of $\bar{\kappa}$, all power allocation strategies generated using the proposed method are to assign power uniformly across subcarriers. Consequently, all the sidelobe levels of the \ac{pacs} are identical, as portrayed in Fig.~\ref{fig:pacs}. As a benchmark, the achievable rate of the ``uniform kurtosis'' strategy assigning identical kurtoses to all subcarriers is also plotted. As can be observed from the figure, the proposed method (which is, in this case, effectively a kurtosis allocation approach) provides a rate gain of up to $0.1$ bits per subcarrier. 

\begin{figure}[t]
    \centering
    \includegraphics[width=0.93\linewidth]{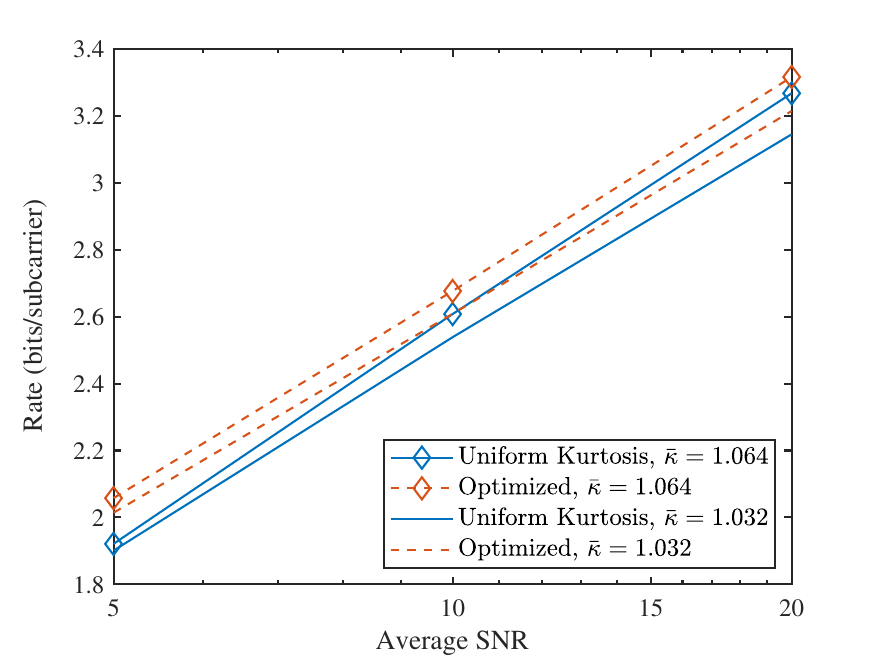}
    \caption{The achievable rate versus the average SNR across subcarriers.}
    \label{fig:rate_vs_snr}
\end{figure}

Next, let us investigate the dependence of the achievable rate on the average SNR over all subcarriers. As can be observed from Fig.~\ref{fig:rate_vs_snr}, the rate gain is larger when the average SNR is smaller, for both $\bar{\kappa}=1.032$ and $\bar{\kappa}=1.064$. This corroborates intuitions since the discrepancy between the channel qualities of different subcarriers diminishes as the average SNR increases.

\begin{figure}[t]
    \centering
    \includegraphics[width=0.93\linewidth]{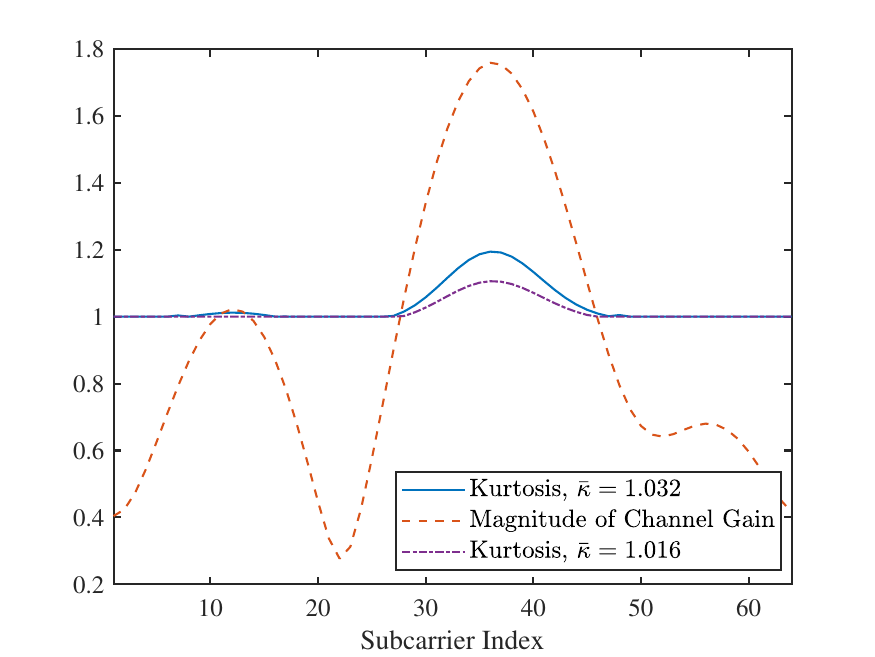}
    \caption{The kurtosis allocation strategy for a given realization of the channel.}
    \label{fig:kurtosis_and_gain}
\end{figure}

For a given channel realization, we plot the channel gains ($|h_i|$'s) as well as the kurtosis allocation vector in Fig.~\ref{fig:kurtosis_and_gain}. Observe that the kurtosis distribution resembles the ``water-filling'' strategy, which assigns the smallest possible kurtosis (namely, $1$) to subcarriers having a channel gain less than a certain threshold. For above-threshold subcarriers, larger kurtoses are assigned to those that have a larger gain. 

\begin{figure}[t]
    \centering
    \includegraphics[width=0.93\linewidth]{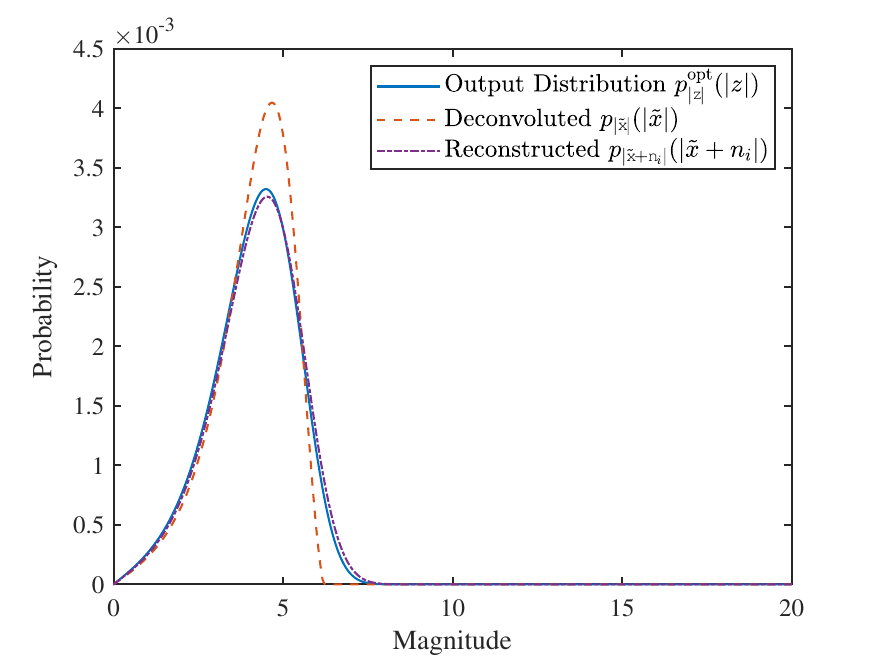}
    \caption{Input distribution obtained by solving the deconvolution problem \eqref{deconvolution_relax}.}
    \label{fig:deconvolution}
\end{figure}

Finally, we demonstrate the input distribution obtained by solving the deconvolution equation \eqref{deconvolution}, with the aid of Fourier transform. In particular, we solve the relaxed deconvolution problem
\begin{subequations}\label{deconvolution_relax}
\begin{align}
\min_{p_{\tilde{\rv{x}}}(\tilde{x})}&~~\left\|\int p_{\tilde{\rv{x}}}(\tilde{x}) p_{\rv{n}_i}(z-\tilde{x}){\rm d}\tilde{x} - p_{\rv{z}}^{\rm opt}(z)\right\|, \label{obj_deconvolution}\\
{\rm s.t.}&~~p_{\tilde{\rv{x}}}(\tilde{x})\geq 0,~\int p_{\tilde{\rv{x}}}(\tilde{x}){\rm d} \tilde{x}=1.
\end{align} 
\end{subequations}
The convolution in \eqref{obj_deconvolution} can be transformed into an ordinary multiplication using the Fourier transform. In the numerical example, we focus on $36$-th subcarrier in the channel realization portrayed in Fig.~\ref{fig:kurtosis_and_gain}. For the convenience of illustration, we only plot the distribution of the magnitudes, which is sufficient since the distribution of phase is always uniform. The output distribution and the input distribution obtained by deconvolution are shown in Fig.~\ref{fig:deconvolution}. We observe that the reconstructed PDF of $p_{|\tilde{\rv{x}}+\rv{n}_i|}(|\tilde{x}+n_i|)$ (obtained by convolving $p_{\tilde{\rv{x}}}(\tilde{x})$ with $p_{\rv{n}_i}(n_i)$ and taking the magnitude) closely approximates the original output distribution $p_{\rv{z}}^{\rm opt}(z)$, implying that the accuracy of the deconvolution is satisfactory.

\section{conclusions}\label{sec:conclusion}
In this treatise, we have proposed a computationally efficient approach to the design of input distributions in \ac{ofdm}-\ac{isac} systems undergoing frequency-selective channels. In particular, the proposed gradient projection-based method relies only on closed-form expressions and one-dimensional searches, and thus may be used in real-time scenarios. Using semi-analytical arguments, we show that the optimal strategy under practical sensing constraints is a uniform power allocation and a water-filling-like kurtosis allocation, which is in stark contrast to pure communication scenarios. Under the proposed framework, future research directions include precoding design and subcarrier selection. Our hope is that this treatise would provide useful insights for practical \ac{isac} system design. 

\section*{ACKNOWLEDGEMENT}
\label{ACKNOWLEDGEMENT}

This work was supported by the National Natural Science Foundation of China (No.~62301060).

\theendnotes

\appendix

\subsection{Proof of Proposition 1}
\label{app:proof_EISL}
Let us first express the \ac{eisl} as
\begin{equation}
{\rm EISL} = \mathbb{E}\left\{\sum_{i=1}^N|[\RV{r}_{\RV{x}}]_i|^2\right\} -\mathbb{E}\{|[\RV{r}_{\RV{x}}]_1|^2\},
\end{equation}
where 
\begin{align}
\mathbb{E}\left\{\sum_{i=1}^N|[\RV{r}_{\RV{x}}]_i|^2\right\}&=\mathbb{E}\{\|\RV{r}_{\RV{x}}\|_2^2\}\nonumber\\
&=\mathbb{E}\{\|\M{F}^{\rm{H}}|\RV{x}|^2\|_2^2\}.
\end{align}
Using Parseval's identity, we obtain
\begin{equation}\label{fourth}
\mathbb{E}\{\|\M{F}^{\rm{H}}|\RV{x}|^2\|_2^2\}=\mathbb{E}\{\||\RV{x}|^2\|_2^2\}=\mathbb{E}\{\|\RV{x}\|_4^4\}.
\end{equation}
For the term $\mathbb{E}\{|[\RV{r}_{\RV{x}}]_1|^2\}$, we have
\begin{align}
\mathbb{E}\{|[\RV{r}_{\RV{x}}]_1|^2\}&=\mathbb{E}\left\{\left|\frac{1}{\sqrt{N}}\V{1}^{\rm T}|\RV{x}|^2\right|^2\right\}\nonumber \\
&=\frac{1}{N}\mathbb{E}\{\|\RV{x}\|_2^4\}.
\end{align}
Using the independence of $\rv{x}_i$'s, we have
\begin{align}
\mathbb{E}\{\|\RV{x}\|_2^4\}&=\mathbb{E}\left(\sum_{i=1}^N\sum_{j=1}^N |\rv{x}_i|^2|\rv{x}_j|^2\right)\nonumber \\
&=\mathbb{E}\{\|\RV{x}\|_4^4\}+(\mathbb{E}\{\|\RV{x}\|_2^2\})^2 \nonumber \\
&\hspace{3mm} - \sum_{i=1}^N (\mathbb{E}\{|\rv{x}_i|^2\})^2.\label{second}
\end{align}
Combining \eqref{fourth} and \eqref{second} yields \eqref{eisl_expression}.

\subsection{Proof of Proposition 2}
\label{app:proof_diff_ent}
Upon denoting $\rv{z}=\rv{x}+i\rv{y}$, we obtain the following expression for $Z_0$
\begin{subequations}
\begin{align}
Z_0&=\int e^{a|z|^2-c|z|^4}{\rm d}z\\
&=\int e^{a(x^2+y^2)-c(x^2+y^2)^2} {\rm d}x{\rm d}y\\
&=\int_0^{2\pi}{\rm d}\theta \int_0^{\infty} \rho e^{a\rho^2-c\rho^4}{\rm d}\rho\\
&=\pi \int_0^{\infty} e^{a\eta-c\eta^2}{\rm d}\eta,
\end{align}
\end{subequations}
where $\rv{\eta} =\rv{\rho}^2$. Next, denote
$$
p(\eta)=\frac{1}{Z_1}e^{a\eta-c\eta^2},
$$
where $\eta \geq 0$ and $Z_1=\int_0^\infty e^{a\eta-c\eta^2}{\rm d}\eta$, and hence $Z_0=\pi Z_1$, we may then write the moments of $\rv{z}$ as follows
\begin{subequations}
\begin{align}
M_2=\mathbb{E}\{|\rv{z}|^2\}&=\int |z|^2p_{\rv{z}}(z){\rm d} z\\
&=\frac{1}{\pi Z_1}\int |z|^2e^{a|z|^2-c|z|^4}{\rm d}z\\
&=\frac{1}{Z_1}\int_0^{\infty}\eta e^{a\eta-c\eta^2}{\rm d}\eta\\
&=\mathbb{E}(\rv{\eta}).
\end{align}
\end{subequations}
Similarly we have $M_4=\mathbb{E}\{|\rv{z}|^4\}=\mathbb{E}(\rv{\eta}^2)$. The differential entropy of $\rv{z}$ takes the form of
\begin{subequations}
\begin{align}
h(\rv{z})&=-\int p_{\rv{z}}(z)\ln p_{\rv{z}}(z){\rm d}z\\
&=-\int \frac{e^{a|z|^2-c|z|^4}}{Z_0} \left(-\ln Z_0+a|z|^2-c|z|^4\right){\rm d}z\\
&=\ln Z_0-a\mathbb{E}\{|\rv{z}|^2\}+c\mathbb{E}\{|\rv{z}|^4\}\\
&=\ln \pi +\ln Z_1 -a\mathbb{E}(\rv{\eta})+c\mathbb{E}(\rv{\eta}^2).
\end{align}
\end{subequations}
Next we rewrite $Z_1$, $\mathbb{E}(\rv{\eta})$ and $\mathbb{E}(\rv{\eta}^2)$ as functions of $a$ and $c$. We commence with $Z_1$:
\begin{subequations}
\begin{align}
Z_1&=\int_0^\infty e^{-c\eta^2+a\eta} {\rm d}\eta\\
&=e^{\frac{a^2}{4c}}\int_0^\infty e^{-c(\eta-\frac{a}{2c})^2}{\rm d}\eta\\
&=e^{\frac{a^2}{4c}}\cdot \frac{1}{\sqrt{c}} \int_{-\frac{a}{2\sqrt{c}}}^\infty e^{-t^2}{\rm d}t \\
&=e^{\frac{a^2}{4c}}\cdot \frac{1}{\sqrt{c}}\cdot \frac{\sqrt{\pi}}{2} \cdot {\rm erfc}\left(-\frac{a}{2\sqrt{c}}\right)\\
&=\sqrt{\frac{\pi}{4c}}e^{\frac{a^2}{4c}}\left[1+{\rm erf}\left(\frac{a}{2\sqrt{c}}\right)\right].
\end{align}
\end{subequations}
Let us now consider $M_2 = \mathbb{E}(\rv{\eta})$
\begin{subequations}
\begin{align}
\mathbb{E}(\rv{\eta})&=\int_0^{\infty} \eta p(\eta){\rm d}\eta\\
&=\frac{1}{Z_1}\int_0^{\infty} \eta (e^{-c\eta^2+a\eta}){\rm d}\eta.
\end{align}
\end{subequations}
Note that
$$
\frac{\mathrm{d}}{\mathrm{d} \eta}(e^{-c\eta^2+a\eta}) = (-2c\eta+a)e^{-c\eta^2+a\eta}.
$$
Integrating both sides, we obtain
$$
\begin{aligned}
\int_0^\infty \frac{\mathrm{d}}{\mathrm{d} \eta}(e^{-c\eta^2+a\eta}){\rm d}\eta &=-2c\int_0^\infty \eta(e^{-c\eta^2+a\eta}){\rm d}\eta\\
&\hspace{3mm}+a\int_0^\infty (e^{-c\eta^2+a\eta}){\rm d}\eta,
\end{aligned}
$$
which amounts to
$$
-1 = -2cZ_1\mathbb{E}(\rv{\eta})+aZ_1,
$$
and hence we have
$$
M_2=\mathbb{E}(\rv{\eta})=\frac{1}{2c}\left(a+\frac{1}{Z_1}\right).
$$
Next we consider $M_4=\mathbb{E}(\rv{\eta}^2)$
\begin{subequations}
\begin{align}
\mathbb{E}(\rv{\eta}^2)&=\int_0^{\infty} \eta^2 p(\eta){\rm d}\eta\\
&=\frac{1}{Z_1}\int_0^{\infty} \eta^2 (e^{-c\eta^2+a\eta}){\rm d}\eta.
\end{align}
\end{subequations}
Similarly, we first compute the derivative
$$
\frac{\mathrm{d}}{\mathrm{d} \eta}(\eta e^{-c\eta^2+a\eta}) = (-2c\eta^2+a\eta+1)e^{-c\eta^2+a\eta},
$$
and then integrate both sides
$$
\begin{aligned}
\int_0^\infty \frac{\mathrm{d}}{\mathrm{d} \eta}(\eta e^{-c\eta^2+a\eta}) {\rm d}\eta &= -2c \int_0^\infty \eta^2 e^{-c\eta^2+a\eta} {\rm d}\eta\\
&\hspace{3mm}+a\int_0^\infty \eta e^{-c\eta^2+a\eta} {\rm d}\eta \\
&\hspace{3mm}+\int_0^\infty e^{-c\eta^2+a\eta} {\rm d}\eta,
\end{aligned}
$$
yielding
$$
0 = -2cZ_1\mathbb{E}(\eta^2)+aZ_1\mathbb{E}(\eta)+Z_1.
$$
Finally we have
$$
M_4 = \mathbb{E}(\rv{\eta}^2) = \frac{1}{2c}\big(1+a\mathbb{E}(\rv{\eta})\big).\\
$$
We may now express the differential entropy of $\rv{z}$ as
\begin{subequations}
\begin{align}
h(\rv{z})&=\ln\pi +\ln Z_1 - aM_2 + cM_4\\
&=\ln\pi +\ln Z_1 - \frac{a}{2c}\left(a+\frac{1}{Z_1}\right)\nonumber\\
&\hspace{3mm}+\frac{1}{2}\left(1+\frac{a^2}{2c}+\frac{a}{2c}\cdot\frac{1}{Z_1}\right)\\
&=\underbrace{\ln \pi +\frac{1}{2}}_{\mathrm{constant}} +\ln Z_1 -\frac{a}{4c}\cdot \frac{1}{Z_1}-\frac{a^2}{4c}.\label{diff_ent_final}
\end{align}
\end{subequations}

\subsection{Proof of Proposition 3}\label{app:proof_grad_z}
From \eqref{diff_ent_final} we see that
\begin{align}
\nabla_{\V{\theta}} h(\rv{z}) &=\left(\frac{1}{Z_1}+\frac{a}{4cZ_1^2}\right)\nabla_{\V{\theta}} Z_1\nonumber \\
&\hspace{3mm}-\frac{1}{4c}\left[2a+\frac{1}{Z_1},-\frac{a}{c}\left(a+\frac{1}{Z_1}\right)\right]^{\rm T}.\label{grad_1}
\end{align}
For $\nabla_{\V{\theta}} Z_1$, we have
\begin{subequations}
\begin{align}
\frac{ \partial Z_1}{ \partial a} &= \frac{ \partial}{\partial a}\int_0^\infty e^{-c\eta^2+a\eta}{\rm d}\eta\\
&=\int_0^\infty \frac{ \partial}{\partial a} e^{-c\eta^2+a\eta}{\rm d}\eta \\
&= \int_0^\infty \eta e^{-c\eta^2+a\eta}{\rm d}\eta\\
&=Z_1M_2, \label{pd_1}
\end{align}
\end{subequations}
and
\begin{subequations}
\begin{align}
\frac{ \partial Z_1}{ \partial c} &= \frac{ \partial}{\partial c}\int_0^\infty e^{-c\eta^2+a\eta}{\rm d}\eta\\
&=\int_0^\infty \frac{ \partial}{\partial c} e^{-c\eta^2+a\eta}{\rm d}\eta \\
&=\int_0^\infty -\eta^2 e^{-c\eta^2+a\eta}{\rm d}\eta\\
&=-Z_1M_4. \label{pd_2}
\end{align}
\end{subequations}
Substituting \eqref{pd_1} and \eqref{pd_2} into \eqref{grad_1}, we obtain
\begin{align}
\frac{ \partial h(\rv{z})}{\partial a} &= \left(\frac{1}{Z_1}+\frac{a}{4cZ_1^2}\right)\frac{\partial Z_1}{ \partial a}-\frac{a}{2c}-\frac{1}{4cZ_1}\nonumber\\
&=\left(\frac{1}{Z_1}+\frac{a}{4cZ_1^2}\right)\left(\frac{aZ_1}{2c}+\frac{1}{2c}\right)-\frac{a}{2c}-\frac{1}{4cZ_1}\nonumber\\
&=\frac{1}{4cZ_1}\left(1+\frac{a^2}{2c}+\frac{a}{2cZ_1}\right)\nonumber\\
&=\frac{1}{2Z_1}M_4,
\end{align}
and
\begin{align}
\frac{ \partial h(z)}{ \partial c} &= \left(\frac{1}{Z_1}+\frac{a}{4cZ_1^2}\right)\frac{ \partial Z_1}{ \partial c}+\frac{a^2}{4c^2}+\frac{a}{4c^2Z_1}\nonumber\\
&=-\frac{1}{2c}-\frac{a}{8c^2Z_1}\left(1+\frac{a^2}{2c}+\frac{a}{2cZ_1}\right)\nonumber\\
&=-\frac{1}{2c}-\frac{a}{4cZ_1}M_4,
\end{align}
which yield \eqref{grad_z}.

\subsection{Proof of Proposition 4}
\label{app:proof_individual_sidelobe}
Using \eqref{pacs}, we obtain
$$
\begin{aligned}
\mathbb{E}\{|[\RV{r}_{\rv{x}}]_i|^2\} &= \V{f}_i^{\rm H}\mathbb{E}\{|\RV{x}|^2(|\RV{x}|^2)^{\rm T}\}\V{f}_i \nonumber \\
&= \V{f}_i^{\rm H}[\V{p}\V{p}^{\rm T}+(\M{K}-\M{I}){\rm diag}(\V{p}\odot\V{p})]\V{f}_i\\
&=|\V{f}_i^{\rm H}\V{p}|^2+\V{f}_i^{\rm H}[(\M{K}-\M{I}){\rm diag}(\V{p}\odot\V{p})]\V{f}_i,
\end{aligned}
$$
where $\M{K}={\rm diag}([\kappa_1,\dotsc,\kappa_N])$. Next, note that
$$
\begin{aligned}
\V{f}_i^{\rm H}[(\M{K}-\M{I}){\rm diag}(\V{p}\odot\V{p})]\V{f}_i= \sum_{j=1}^N |[\V{f}_i]_j|^2 (\kappa_i-1)p_i^2,
\end{aligned}
$$
and that all entries of $\M{F}$ have identical magnitudes, we arrive at
$$
\V{f}_i^{\rm H}[(\M{K}-\M{I}){\rm diag}(\V{p}\odot\V{p})]\V{f}_i= \frac{1}{N}\sum_{j=1}^N (\kappa_i-1)p_i^2,
$$
which yields \eqref{exp_pacs}.

\bibliographystyle{gbt7714-numerical}
\bibliography{references,references_SPM,database}

@article{du2024reshaping,
  title={Reshaping the {ISAC} Tradeoff Under {OFDM} Signaling: A Probabilistic Constellation Shaping Approach},
  author={Du, Zhen and Liu, Fan and Xiong, Yifeng and others},
  journal={IEEE Transactions on Signal Processing},
  year={2024},
  volume={72},
  number={},
  pages={4782-4797}
}

@ARTICLE{9724170,
  author={Ye, Zhifan and Zhou, Zhengchun and Fan, Pingzhi and others},
  journal={IEEE Journal on Selected Areas in Communications}, 
  title={Low Ambiguity Zone: {T}heoretical Bounds and {D}oppler-Resilient Sequence Design in Integrated Sensing and Communication Systems}, 
  year={2022},
  volume={40},
  number={6},
  pages={1809-1822},
  doi={10.1109/JSAC.2022.3155510}}

@ARTICLE{liao2024pulse,
  author={Liao, Zihan and Liu, Fan and Li, Shuangyang and others},
  journal={IEEE Transactions on Wireless Communications}, 
  title={Pulse Shaping for Random {ISAC} Signals: {T}he Ambiguity Function Between Symbols Matters}, 
  year={2025},
  volume={24},
  number={4},
  pages={2832-2846},
  doi={10.1109/TWC.2024.3525440}}

@ARTICLE{zhang2023input,
  author={Zhang, Yumeng and Aditya, Sundar and Clerckx, Bruno},
  journal={IEEE Transactions on Signal Processing}, 
  title={Input Distribution Optimization in {OFDM} Dual-Function Radar-Communication Systems}, 
  year={2024},
  volume={72},
  number={},
  pages={5258-5273},
  doi={10.1109/TSP.2024.3491899}}

@ARTICLE{Keskin2024fundamental,
  author={Keskin, Musa Furkan and Mojahedian, Mohammad Mahdi and Lacruz, Jesus O. and others},
  journal={IEEE Transactions on Wireless Communications}, 
  title={Fundamental Trade-Offs in Monostatic {ISAC}: {A} Holistic Investigation Towards {6G}}, 
  note={\textit{early access.}},
  year={2025},
  volume={},
  number={},
  pages={},
  doi={10.1109/TWC.2025.3563197}}

@article{1,
  title={Wireless Communications-OFDM and Its Wireless Applications: A Survey},
  author={Hwang, Taewon and Yang, Chenyang and Wu, Gang and Li, Shaoqian and Li, GY},
  journal={IEEE Transactions on Vehicular Technology},
  volume={58},
  number={4},
  pages={1673},
  year={2009}
}

@article{2,
  title={LTE-advanced: next-generation wireless broadband technology},
  author={Ghosh, Amitava and Ratasuk, Rapeepat and Mondal, Bishwarup and Mangalvedhe, Nitin and Thomas, Tim},
  journal={IEEE wireless communications},
  volume={17},
  number={3},
  pages={10--22},
  year={2010},
  publisher={IEEE}
}

@article{20,
  title={Peak-to-average power ratio reduction of OFDM signals with nonlinear companding scheme},
  author={Hou, Jun and Ge, Jianhua and Zhai, Dewei and Li, Jing},
  journal={IEEE Transactions on Broadcasting},
  volume={56},
  number={2},
  pages={258--262},
  year={2010},
  publisher={IEEE}
}

@inproceedings{liu2023deterministic,
  title={Deterministic-random tradeoff of integrated sensing and communications in Gaussian channels: {A} rate-distortion perspective},
  author={Liu, Fan and Xiong, Yifeng and Wan, Kai and Han, Tony Xiao and Caire, Giuseppe},
  booktitle={Proc. 2023 IEEE International Symposium on Information Theory (ISIT)},
  pages={2326--2331},
  year={2023},
  address={Taipei, China},
  organization={IEEE}
}

@ARTICLE{has_isac,
  author={Li, Xinyu and Cui, Yuanhao and Zhang, J. Andrew and Liu, Fan and Zhang, Daqing and Hanzo, Lajos},
  journal={IEEE Communication Magazine}, 
  title={Integrated Human Activity Sensing and Communications}, 
  year={2023},
  volume={61},
  number={5},
  pages={90-96},
  doi={10.1109/MCOM.002.2200391}}

@article{preprint_opt_ofdm,
  title={{OFDM} achieves the lowest ranging sidelobe under random {ISAC} signaling},
  author={Liu, Fan and Zhang, Ying and Xiong, Yifeng and Li, Shuangyang and Yuan, Weijie and Gao, Feifei and Jin, Shi and Caire, Giuseppe},
  journal={arXiv preprint arXiv:2407.06691},
  year={2024},
  url={https://arXiv.org/abs/2407.06691}
}

@ARTICLE{9787809,
  author={Liu, Yao and Li, Min and Liu, An and Lu, Jianmin and Han, Tony Xiao},
  journal={IEEE Transactions on Vehicular Technology}, 
  title={Information-Theoretic Limits of Integrated Sensing and Communication With Correlated Sensing and Channel States for Vehicular Networks}, 
  year={2022},
  volume={71},
  number={9},
  pages={10161-10166},
  doi={10.1109/TVT.2022.3179869}}

@ARTICLE{10153971,
  author={Ahmadipour, Mehrasa and Wigger, Mich\`{e}le},
  journal={IEEE Journal on Selected Areas in Information Theory}, 
  title={An Information-Theoretic Approach to Collaborative Integrated Sensing and Communication for Two-Transmitter Systems}, 
  year={2023},
  volume={4},
  number={},
  pages={112-127},
  doi={10.1109/JSAIT.2023.3286932}}

@ARTICLE{iceberg,
  author={Liu, Fan and Xiong, Yifeng and Lu, Shihang and Li, Shuangyang and Yuan, Weijie and Masouros, Christos and Jin, Shi and Caire, Giuseppe},
  journal={IEEE Transactions on Signal Processing}, 
  title={Uncovering the Iceberg in the Sea: {F}undamentals of Pulse Shaping and Modulation Design for Random {ISAC} Signals}, 
  year={2025},
  volume={73},
  number={},
  pages={2511-2526},
  doi={10.1109/TSP.2025.3580596}}

@techreport{ran_meeting,
 author = {3GPP},
 institution = {{3rd Generation Partnership Project (3GPP)}},
 title = {{New SID: Study on 6G Radio}},
 type = {Work Item Description (WID)},
 year = {2025}
}

@ARTICLE{Chafii2023CST,
  author={Chafii, Marwa and Bariah, Lina and Muhaidat, Sami and Debbah, Merouane},
  journal={IEEE Communications Surveys and Tutorials}, 
  title={Twelve Scientific Challenges for {6G: Rethinking} the Foundations of Communications Theory}, 
  month={Secondquarter},
  year={2023},
  volume={25},
  number={2},
  pages={868-904},
  doi={10.1109/COMST.2023.3243918}}

@article{saad2019vision,
  title={A vision of {6G} wireless systems: {A}pplications, trends, technologies, and open research problems},
  author={Saad, Walid and Bennis, Mehdi and Chen, Mingzhe},
  journal={IEEE Network},
  volume={34},
  number={3},
  pages={134--142},
  year={2019},
  publisher={IEEE}
}

@article{ITU2023,
  title={{D}raft {N}ew {Recommendation ITU-R M. [IMT. Framework
for 2030 and Beyond]}},
  author={{ITU-R WP5D}},
  year={2023}
}

@ARTICLE{9921271,
  author={Wei, Zhiqing and Wang, Yuan and Ma, Liang and others},
  journal={IEEE Transactions on Vehicular Technology}, 
  title={{5G PRS}-Based Sensing: {A} Sensing Reference Signal Approach for Joint Sensing and Communication System}, 
  year={2023},
  volume={72},
  number={3},
  pages={3250-3263},
  doi={10.1109/TVT.2022.3215159}}

@ARTICLE{10471902,
  author={Xiong, Yifeng and Liu, Fan and Wan, Kai and Yuan, Weijie and Cui, Yuanhao and Caire, Giuseppe},
  journal={IEEE BITS the Information Theory Magazine}, 
  title={From Torch to Projector: {F}undamental Tradeoff of Integrated Sensing and Communications}, 
  year={2024},
  volume={},
  number={},
  pages={1-13},
  doi={10.1109/MBITS.2024.3376638}}

@ARTICLE{10147248,
  author={Xiong, Yifeng and Liu, Fan and Cui, Yuanhao and others},
  journal={IEEE Transactions on Information Theory}, 
  title={On the Fundamental Tradeoff of Integrated Sensing and Communications Under {G}aussian Channels}, 
  year={2023},
  volume={69},
  number={9},
  pages={5723-5751},
  doi={10.1109/TIT.2023.3284449}}

@ARTICLE{9785593,
  author={Ahmadipour, Mehrasa and Kobayashi, Mari and Wigger, Michele and Caire, Giuseppe},
  journal={IEEE Transactions on Information Theory}, 
  title={An Information-Theoretic Approach to Joint Sensing and Communication}, 
  year={2022},
  volume={70},
  number={2},
  pages={1124-1146}
}

@ARTICLE{10638525,
  author={Yuan, Weijie and Zhou, Lin and Dehkordi, Saeid K. and others},
  journal={IEEE Wireless Communications}, 
  title={From {OTFS} to {DD-ISAC}: {I}ntegrating Sensing and Communications in the Delay {Doppler} Domain}, 
  year={2024},
  volume={31},
  number={6},
  pages={152-160},
  doi={10.1109/MWC.018.2300607}}

@ARTICLE{9737357,
  author={Liu, Fan and Cui, Yuanhao and Masouros, Christos and others},
  journal={IEEE Journal on Selected Areas in Communications}, 
  title={Integrated Sensing and Communications: {T}oward Dual-Functional Wireless Networks for 6{G} and Beyond}, 
  year={2022},
  volume={40},
  number={6},
  pages={1728-1767},
  doi={10.1109/JSAC.2022.3156632}}

@article{sturm2011waveform,
  title={Waveform Design and Signal Processing Aspects for Fusion of Wireless Communications and Radar Sensing},
  author={C. Sturm and W. Wiesbeck},
  journal={Proceedings of IEEE},
  volume={99},
  number={7},
  pages={1236-1259},
  year={2011},
  month={Jul.} ,
}

@ARTICLE{9359665,
  author={Chen, Xu and Feng, Zhiyong and Wei, Zhiqing and Zhang, Ping and Yuan, Xin},
  journal={IEEE Internet Things Journal}, 
  title={Code-Division {OFDM} Joint Communication and Sensing System for 6{G} Machine-Type Communication}, 
  year={2021},
  volume={8},
  number={15},
  pages={12093-12105},
  doi={10.1109/JIOT.2021.3060858}}

@ARTICLE{9005192,
  author={Zeng, Yonghong and Ma, Yugang and Sun, Sumei},
  journal={IEEE Transactions on Vehicular Technology}, 
  title={Joint Radar-Communication With Cyclic Prefixed Single Carrier Waveforms}, 
  year={2020},
  volume={69},
  number={4},
  pages={4069-4079},
  doi={10.1109/TVT.2020.2975243}}

@ARTICLE{9109735,
  author={Gaudio, Lorenzo and Kobayashi, Mari and Caire, Giuseppe and Colavolpe, Giulio},
  journal={IEEE Transactions on Wireless Communications}, 
  title={On the Effectiveness of {OTFS} for Joint Radar Parameter Estimation and Communication}, 
  year={2020},
  volume={19},
  number={9},
  pages={5951-5965},
  doi={10.1109/TWC.2020.2998583}}

@ARTICLE{10264814,
  author={Tagliaferri, Dario and Mizmizi, Marouan and Mura, Silvia and others},
  journal={IEEE Transactions on Wireless Communications}, 
  title={Integrated Sensing and Communication System via Dual-Domain Waveform Superposition}, 
  year={2024},
  volume={23},
  number={5},
  pages={4284-4299},
  doi={10.1109/TWC.2023.3316888}}

@ARTICLE{10463758,
  author={Keskin, Musa Furkan and Marcus, Carina and Eriksson, Olof and others},
  journal={IEEE Transactions on Wireless Communications}, 
  title={Integrated Sensing and Communications with {MIMO-OTFS: ISI/ICI} Exploitation and Delay-{D}oppler Multiplexing}, 
  year={2024},
  volume={23},
  number={8},
  pages={10229-10246},
  doi={10.1109/TWC.2024.3370501}}

@ARTICLE{10012421,
  author={Wei, Zhiqing and Qu, Hanyang and Wang, Yuan and others},
  journal={IEEE Internet Things Journal}, 
  title={Integrated Sensing and Communication Signals Toward {5G-A and 6G: A} Survey}, 
  year={2023},
  volume={10},
  number={13},
  pages={11068-11092},
  doi={10.1109/JIOT.2023.3235618}}

\biographies

\begin{CCJNLbiography}{photo_weijiang.jpg}{Weijiang Zhao}
received the B.S. degree from Beijing University of Posts and Telecommunications (BUPT) in 2024. He is currently pursuing his PhD degree in the School of Information and Communication Engineering, BUPT. His current research interests include integrated sensing and communications, wireless communication theory and technology.
\end{CCJNLbiography}

\begin{CCJNLbiography}{yifeng_photo_type2.jpg}{Yifeng Xiong}
received the B.S. and M.S. degree from Beijing Institute of Technology in 2015 and 2018, respectively, and the PhD degree from University of Southampton in 2022. He is currently an Associate Professor with the School of Information and Communication Engineering, Beijing University of Posts and Telecommunications (BUPT). His research interests include integrated sensing and communications, quantum computation, quantum information theory, and statistical inference over networks. He was a recipient of the 2025 IEEE Communication Society \& Information Theory Society Joint Paper Award, the Best Master Thesis Award of Chinese Institute of Electronics, and the Best Paper Award of IEEE/CIC ICCC 2023.
\end{CCJNLbiography}

\end{document}